\documentclass[reprint,amsmath,amssymb]{revtex4-1}

\usepackage{graphicx}
\usepackage{amsmath}
\usepackage{dcolumn}
\usepackage{bm}
\usepackage{xcolor}
\usepackage[utf8x]{inputenc}
\usepackage{hyperref}
\begin{document}

\title{Laser-Pulse and Electron-Bunch Plasma Wakefield Accelerator}%

\author{Tianhong Wang$^1$, Vladimir  Khudik$^1$$^,$$^2$, and Gennady Shvets$^1$}
 \affiliation{$^1$School of Applied and Engineering Physics, Cornell University, Ithaca, New York 14850, USA.\\$^2$Department of Physics and Institute for Fusion Studies, The University of Texas at Austin, Austin, Texas 78712, USA.}

\date{\today}%

\begin{abstract}
Propagation distances of intense laser pulses and high-charge electron beams through the plasma are, respectively, limited by diffraction and self-deceleration. This imposes severe constraints on the performance of the two major advanced accelerator concepts: laser and plasma wakefield accelerators. Using numerical simulations, we demonstrate that when the two beams co-propagate in the plasma, they can interact synergistically and extend each other's travel distances. The key interactions responsible for the synergy are found to be laser channeling by the electron bunch, and direct laser acceleration of the bunch electrons by the laser pulse. Remarkably, the amount of energy transferred from the laser pulse to the plasma can be increased by several times by the guiding electron bunch despite its small energy content. Implications of such synergistic interactions for the high-gradient acceleration of externally injected witness charges are discussed, and a new concept of a Laser-pulse and Electron-bunch Plasma Accelerator (LEPA) is formulated.
\end{abstract}

\maketitle


\section{Introduction}

Plasma-based accelerators represent one of the most exciting concepts in high-gradient particle acceleration. Plasmas can sustain high accelerating gradients on the order of tens to hundreds of ${\rm GV/m}$, thereby enabling compact particle accelerators that are much smaller than the present-day conventional accelerators. The two major approaches to plasma-based acceleration are defined by the way plasma waves are excited: either by relativistic electron bunches for a plasma wakefield accelerator (PWFA)~\cite{PWFA_1}, or by ultra-intense laser pulses for a laser wakefield accelerator (LWFA)~\cite{LWFA_1,LWFA_2,LWFA_3}. Recent advances in laser technologies further contributed to the remarkable successes of the LWFA scheme: generation of low-emittance multi-GeV electron beams have been produced using petawatt-scale laser systems around the world~\cite{GeV_0,GeV_1,GeV_2,GeV_3,GeV_2b}. In addition to their potential role in developing TeV-scale linear lepton colliders~\cite{Colliders}, LWFAs will likely contribute to a wide range of application, such as compact X-ray radiation sources~\cite{Park2006,Kneip2008, Stark} and novel sources of other energetic particles: ions~\cite{Schollmeier2015}, neutrons~\cite{Pomerantz2014}, and positrons~\cite{Positron_Gahn,Positron_Sarri,Positron_Xu,Positron_Alejo}.

LWFA and PWFA concepts have their unique advantages and limitations. Those are determined by the two factors limiting the single-stage energy gain of the accelerated electrons: (i) the accelerating gradient $E_{\parallel}$, which scales with the plasma density $n_0$ according to $E_{\parallel}\propto n_0^{1/2}$, where the proportionality coefficient is determined by the strength of the driver, and (ii) the acceleration distance $L_{\rm acc}$, which is subject to very different constraints for the LWFA and PWFA concepts. If the strength of either driver is sufficiently high to expel plasma from its path, then the accelerating gradient can be estimated as $E_{\parallel} \sim \sqrt{n_0/10^{18}{\rm cm^{-3}}} [{\rm GV/cm}]]$.

On the other hand, comparing $L_{\rm acc}$ for the two drivers is less straightforward. For a LWFA, $L_{\rm acc}$ is the smallest among the propagation distance $L_{\rm prop}^{\rm laser}$ and the dephasing distance $L_d\propto n_0^{-3/2}$~\cite{Tajima_1979,Joshi_1984,Lu_GeV} between the accelerated electrons and plasma wave. Therefore, assuming that laser diffraction and depletion can be overcome (i.e. $L_{\rm prop}^{\rm laser} > L_d$), it is advantageous to decrease the plasma density in order to maximize the energy gain $\Delta W = E_{\parallel} L_d \propto n_0^{-1}$.  Lower plasma densities $n_0 \sim 10^{17}{\rm cm^{-3}}$ employed in recent experiments ~\cite{GeV_2b} are almost two orders of magnitude less dense than in some of the earlier work~\cite{Nature_10e9}. However, reducing the plasma density presents challenges to maintaining laser guiding over such long distances (tens of centimeters) without diffraction.

While plasma "bubbles" produced by the ponderomotive pressure of an intense laser pulse can be used to overcome diffraction using the phenomenon of relativistic self-guiding, the latter requires that the laser power $P$ significantly exceeds the critical power in the plasma, $P_{\rm crit} = 17(\omega_0/\omega_p)^2 {\rm GW}$~\cite{LWFA_2}, where $\omega_0$ and $\omega_p = \sqrt{4\pi e^2n_0/m}$ are the laser frequency and plasma frequencies, $-e$ and $m$ are the electron charge and mass, respectively. For example, $P > 20P_{\rm crit}$ was found to be optimal for $n_0 \sim10^{17}{\rm cm^{-3}}$~\cite{Lu_GeV,Ibbotson_Pcrit}. In order to achieve the energy gain of $\Delta W \sim 10 {\rm GeV}$ for such tenuous plasmas, $P_L \approx 3 {\rm PW}$ is required for a $\lambda_0 \equiv 2\pi c/\omega_0 = 1 {\rm \mu m}$ laser pulse. For the optimal laser pulse duration $\tau_L \sim \lambda_p/2c$ (where $\lambda_p \equiv 2\pi c/\omega_p$ is the plasma wavelength and $c$ is the speed of light), the total laser energy $U^{\rm laser}$ scales as $U^{\rm laser} \propto \lambda_p^3 \propto n_0^{-3/2}$ with plasma density. This presents an additional challenge for tenuous plasmas, as the pulse energy must be increased to tens of joules. Even though preformed plasma channels can improve laser guiding~\cite{GeV_2,GeV_2b}, their advantage is manifested during the final segments of laser propagation, i.e. after the laser pulse is too depleted to produce its own plasma bubble. Moreover, uncertainties associated with the hydrodynamic process of plasma expansion on a nanosecond scale would make the plasma channel less predictable. And the non-uniform density of the plasma ions produced, for example, by a preceding "heater" pulse~\cite{Milchberg_heater_1993,Milchberg_heater_1995,Milchberg_heater_1999} produces a nonlinear focusing force, and can be deleterious to the emittance of the accelerated electrons.

In contrast, the propagation distance $L_{\rm prop}^{\rm bunch}$ of an electron bunch driver with density $n_b \gg n_0$ is not limited by transverse beam spreading because it experiences linear ion focusing inside the self-generated plasma bubble. The bunch charge $q$ required for generating a fully-evacuated bubble can be estimated as $q=4\pi e \bar{Q} c^3 n_0/\omega_p^3$, where $\bar{Q}>1$ is the normalized bunch charge~\cite{stupakov_2016,My_Driver_2017}. For example, for $n_0 = 10^{17}$cm$^{-3}$ ($c/\omega_p \approx 16.8 {\rm \mu m}$)and $\bar{Q} = 1$, the required electron charge is $q \approx 0.95$nC. Assuming that the energy of a bunch electron is $\gamma_b mc^2 = 0.5$GeV, the total energy of such a bunch is a very modest $U^{\rm bunch} = (q/e) \gamma_b mc^2 \approx 0.475$J. Therefore, it may appear that the plasma density scaling of the bunch charge required to drive a strong plasma wave is more favorable for low-density regimes than the scaling of the corresponding laser pulse energy: $q^{\rm bunch} \propto n_0^{-1/2}$ versus $U^{\rm laser} \propto n_0^{-3/2}$. Nevertheless, the inherent limitation of a PWFA scheme is that the transformer ratio of a high-current driver bunch is severely limited by the extremely strong self-generated decelerating electric field, which is typically on the same order as the peak accelerating field inside the bubble. According to the transformer ratio theorem~\cite{TR_PisinChen_PRL86}, the maximum energy gain of an accelerated (witness) electron beam is limited to $\Delta W_{\rm PWFA} = 2\gamma_b mc^2$.

Thus, it would be highly advantageous to find a way of combining the LWFA and PWFA approaches to benefit from their respective advantages: long propagation distance $L_{\rm prop}^{\rm laser}$ of a laser pulse and a small energy content $U^{\rm bunch}$ of an electron driver bunch. The synergy between the two schemes could be realized if (i) some of the large energy content $U^{\rm laser}$ of the laser pulse could be expended to extend the relative short propagation distance $L_{\rm prop}^{\rm bunch}$, and (ii) the self-guided electron bunch could be used to further extend the laser propagation length. In the following, we show that objective (i) can be accomplished using the recently discovered phenomenon of direct laser acceleration (DLA) in a decelerating plasma wakefield~\cite{Farfield}, and objective (ii) can be accomplished via beam-channeling of laser pulses by high-current electron bunches~\cite{BeamChannel_Shvets_PhysRevE97}.

\begin{figure}[htp!]
\centering
  \includegraphics[width=0.8\columnwidth]{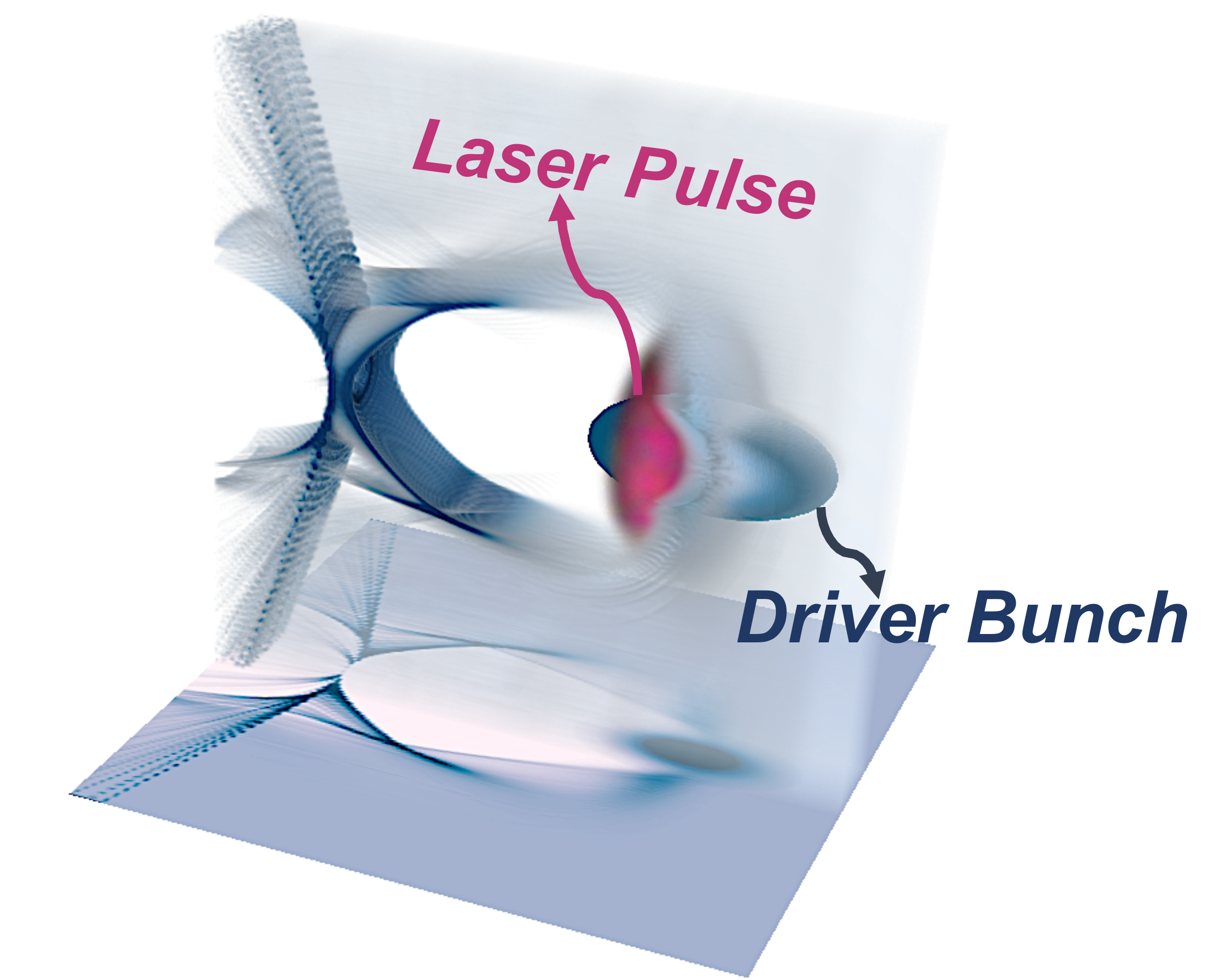}\\
\caption{Schematic of a Laser-pulse and Electron-bunch Plasma Accelerator (LEPA): the bunch guides the laser pulse, and the laser pulse extends the propagation distance of the bunch via direct laser acceleration.}
\label{Fig1}
\end{figure}

The schematic of the combined Laser-pulse and Electron-bunch Plasma Accelerator (LEPA) is shown in Fig.~\ref{Fig1}.  The laser pulse and the electron driver bunch in LEPA are temporally overlapped and are traveling in the same direction. The electron bunch creates a deep plasma bubble using just a fraction of the laser energy ($U^{\rm bunch} \ll U^{\rm laser}$), guides the laser pulse propagation, and mitigates its diffraction. The condition for channeling a laser pulse by an electron bunch has been estimated as $I > I_A/4$, where $I$ is the bunch current and $I_A = mc^3/e \approx 17{\rm kA}$ is the Alfven current. Assuming that the bunch duration $\tau^{\rm bunch} \sim \pi \omega_p^{-1}$ and $I=q/\tau^{\rm bunch}$, we find that the channeling condition is simplified to $\bar{Q} > 1$. Note that the same condition must be satisfied to produce a fully-evacuated plasma bubble.

At the same time, relativistic electrons of the bunch traveling inside the plasma bubble gain energy from the laser through the DLA mechanism~\cite{Shaw_PPCF14,Xi_prl,Zhang_PPCF16,Shaw_PPCF16,Universal}, thereby overcoming deceleration by the self-generated plasma wakefield~\cite{Farfield}. Ideally, driver electrons can gain energy from the laser at twice the rate of their energy loss to the wakefield~\cite{Farfield}: $d \gamma/dt \approx a_{\parallel}\omega_0$, where $a_\parallel = eE_{\parallel}/m\omega_0c \ll 1$ is the normalized longitudinal wakefield. According to a simplified theoretical estimate~\cite{Farfield}, the driver bunch electrons can gain energy up to $\gamma_{\rm max} \approx (\omega_p/\omega_0)^2 (a_0/a_{\parallel})^4/130$ from a laser pulse with a normalized vector potential $a_0 = eE_0/m\omega_0c$, where $E_0$ is the amplitude of the laser electric field. Assuming the wakefield equals to the cold plasma wave-breaking limit, i.e. $a_\parallel = a_\parallel^{\rm WB} \equiv \omega_p/\omega_0$, we estimate that the driver bunch electrons can gain up to several GeVs of energy in a tenuous plasma with $n_0 = 10^{17}{\rm cm^{-3}}$. Therefore, the propagation distance of the electron driver bunch can be indeed extended by the laser pulse.

\section{LEPA parameters, modeling, and key conclusions}\label{sec:params_sims}

\subsection{Parameters selection}\label{sec:params}

While the parameter space for the driver bunch and the laser pulse is very wide, we will concentrate on exploring multi-GeV acceleration using sub-PW laser pulses with $\lambda_0 = 0.8 {\rm \mu m}$ and sub-GeV driver bunches. Multi-GeV acceleration requires tenuous plasma. Therefore, we chose $n_0=4\times10^{17}$ cm$^{-3}$ corresponding to $c/\omega_p \approx 8.4 {\rm \mu m}$ and $P_{\rm crit} \approx 75$TW. To ensure self-focusing, the laser power is chosen to be $P=380$TW. Further, the Gaussian spot size and duration are chosen to be $w=29{\rm \mu m}$ and $\tau_{FWHM}=68$fs, respectively. These laser parameters corresponds to the normalized vector potential $a_0=3.7$.

The charge $Q_b=1.25$nC of a Gaussian electron driver bunch was chosen to correspond to the normalized charge $\bar{Q} \approx 2.6$~\cite{My_Driver_2017} to produce a fully-formed plasma bubble with $a_\parallel > a_\parallel^{\rm WB}$. The transverse size $w_b=8.4\mu$m and duration $\tau_b=56$fs of the bunch were chosen to approximately match $w_b \sim c/\omega_p$ and $\tau_b \sim 2\omega_p^{-1}$, resulting in the peak current $I_{b}=22.3$kA: sufficient to channel the laser pulse because $I_b > I_A$ ~\cite{BeamChannel_Shvets_PhysRevE97}. Driver electrons started with the initial energy $\gamma_b mc^2 = 0.65 {\rm GeV}$ and the energy spread of $0.5{\rm MeV}$. Therefore, the total energy of the electron bunch is $U^{\rm bunch} \approx 0.8{\rm J}$, i.e. just $3\%$ of the laser pulse energy   $U^{\rm laser} \approx 27 {\rm J}$.

\subsection{Computational approach}

Simulating centimeters-long propagation of both laser pulse and driver bunch using conventional particle-in-cell (PIC) codes presents a unique computational challenge because of the high temporal resolution requirements imposed by the DLA mechanism and because of the equally high spatial resolution requirements imposed by the Courant stability condition. In order to capture the laser-electron interactions with sufficient accuracy, a longitudinal spatial step $\Delta z$ and time step $c\Delta t$ close to $\lambda_L/50\sim\lambda_L/100$ are needed~\cite{DLA_require_Alex,My_DLA_2019}.
Therefore, we have selected to carry out fully three-dimensional quasi-static particle-in-cell (QPIC) simulations using an in-house developed code WAND-PIC~\cite{WAND_PIC}. WAND-PIC uses a quasi-static approach~\cite{Mora_1997,Lotov_PiC_2003,QuickPiC_2006,HiPACE_2014,My_Driver_2017} to model the motion of the bubble-forming plasma particles and of the laser pulse envelope (amplitude and phase). This approach reduces the dimensionality of the simulation by one. It also reduces the required longitudinal resolution ($\xi = z - ct$) in the moving reference frame to $\delta \xi \sim c/\omega_p$ scale. Although the longitudinal resolution has been reduced, the nonlinearity of the wakefield, especially in the back of the bubble, is captured by using adaptive grid size in $\xi$ direction, i.e., the step size $\delta\xi$ is adjusted at every step based on the fastest longitudinal velocity of the plasma particles. On the other hand, full equations of motion are used to model the interaction between driver electrons and the high-frequency laser fields. The code accurately models resonant interactions between the bunch electrons executing betatron undulations and the laser pulse using the sub-cycling method,  (see the comparison of WAND-PIC and a full-PIC code in~\cite{My_DLA_2019}). Such interactions are at the heart of the DLA, and their extreme sensitivity to the phase velocity of the laser field necessitates high temporal resolution.

\begin{figure}[htp!]
\centering
  \includegraphics[width=0.99\columnwidth]{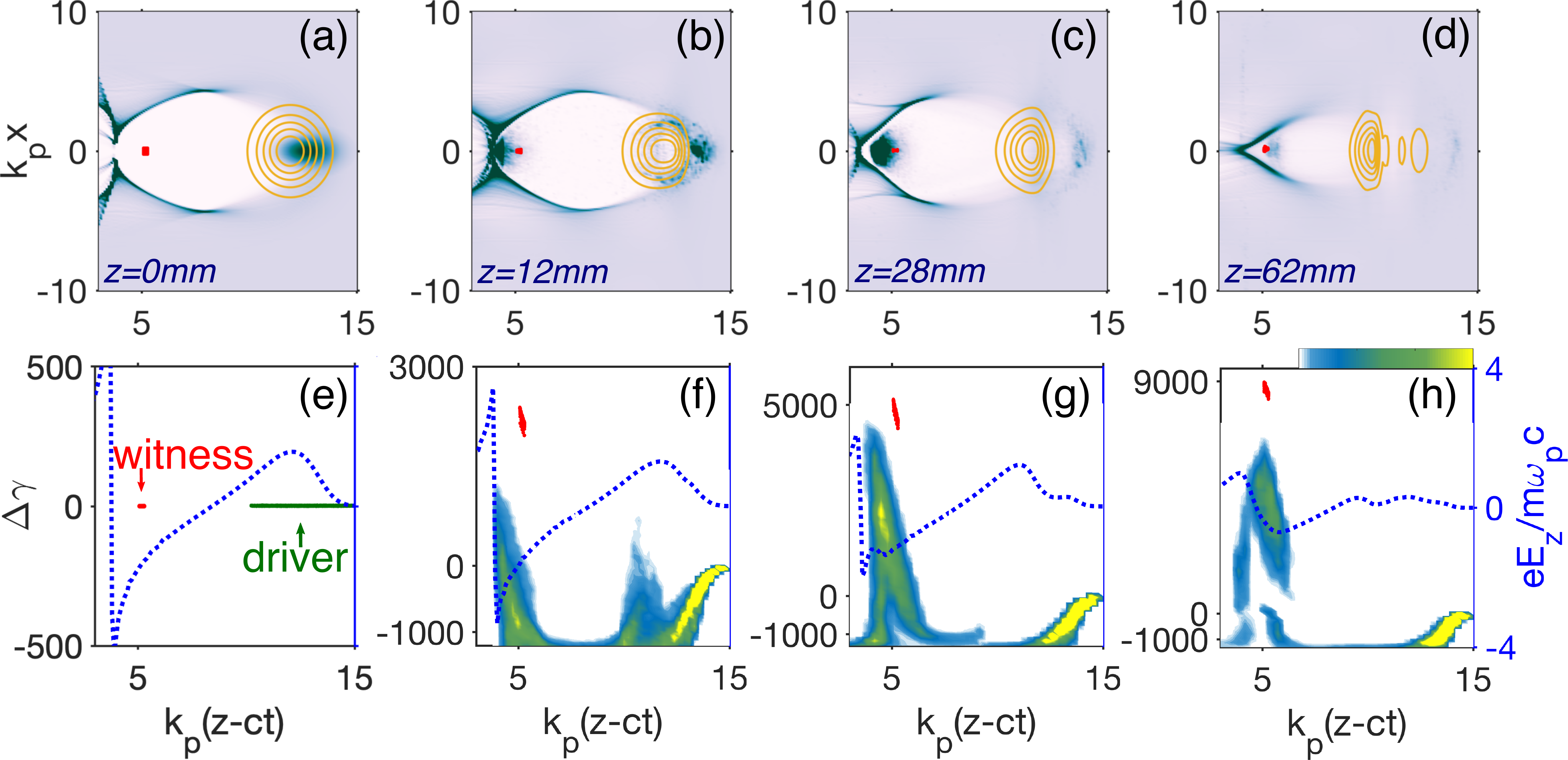}\\
\caption{Evolution of plasma bubble (top row) and the phase space of the witness (red dots) and driver bunch (colormap) electrons (bottom row). (a-d) The $x-z$ cross-section of the plasma bubble at propagation distances $z_0=0$mm, $z_1=12$mm, $z_2 = 28$mm, and $z_3 = 62$mm, respectively. Yellow lines: laser intensity contours, red dots: witness electrons. (e-h) Longitudinal phase space $(\xi=z-ct,\Delta\gamma)$ of the witness and driver bunch electrons, colormaps represent the densities of driver electrons in phase space. Blue dashed lines: longitudinal wakefield $E_z$. Witness beam parameters: $s_x\times s_y\times s_z=4{\rm \mu m}\times 4{\rm \mu m}\times 2{\rm \mu m}$. Beam loading by the witness bunch is neglected in this simulation. Laser-plasma parameters: laser wavelength $\lambda_0 = 0.8\mu$m, normalized laser potential $a_0=3.7$, spot size $w=29{\rm \mu m}$, pulse duration $\tau_{FWHM}=68$fs, plasma density $n_0=4\times10^{17}{\rm cm^{-3}}$. Bunch parameters: charge $Q_b=1.25$nC, transverse size $w_b=8.4\mu$m, duration $\tau_b=56$fs, and energy $\gamma_b mc^2 = 0.65 {\rm GeV}$. }
\label{Fig2}
\end{figure}

\subsection{Acceleration stages of LEPA}

The key results of our simulations are shown in Fig.~\ref{Fig2}. The top row shows the color-coded electron density of the plasma at the propagation distances $z=0$mm, $12$mm, $28$mm, and $62$mm. The plasma bubble is clearly defined. The witness electron bunch is placed near the back of the bubble, where the accelerating field $E_{\parallel}=E_z$ (see Figs.~\ref{Fig2}~(e-h), blue-dotted line) is large. The longitudinal $(\xi, \Delta\gamma)$ phase spaces of the driver (dark-green dots) and witness (red dots) electrons are presented in Figs.~\ref{Fig2}~(e-h), where $\Delta\gamma$ is the change of the relativistic Lorentz factor and $\xi=z-ct$ is the longitudinal position in the moving frame. In this specific simulation, the initial energy of the witness beam is taken to be $\gamma_w mc^2 = 50$MeV. This particular choice is unimportant because future TeV-scale plasma-based accelerators will be segmented into at least a hundred independently-driven segments (stages). The initial energy $\gamma_w mc^2$ will depend on the segment number. Therefore, for all but the first acceleration segment, the following relationship between the energies of the driver and witness beams will be assumed: $\gamma_w \gg \gamma_b$. Therefore, we will be concentrating on the energy {\it gain} $(\Delta\gamma)$ of the witness beam, not on its absolute energy.

Below we introduce the three conceptual stages of LEPA that are differentiated from each other by the importance of the DLA for the driver bunch propagation, as well as the role played by the driver bunch in channeling the laser pulse and generating the plasma bubble. The first stage starts immediately at $z=0$. During this stage, the plasma bubble is produced mainly by the driver bunch, and its size is somewhat larger than that of the bubble produced by the laser pulse alone. Therefore, a higher accelerating gradient is generated in the back (accelerating) portion of the bubble. The first stage of LEPA (from $z_0 = 0$mm to $z_1 =12$mm for this specific example) is characterized by highly efficient DLA: almost $50\%$ of the electrons from the driver bunch experience significant DLA. For example, we observe from Fig.~\ref{Fig2}~(f) that some of the driver bunch electrons in the front portion of the bubble ($(z-ct)>10k_p^{-1}$) have gained $\sim500{\rm MeV}$ of energy. The bunch-averaged DLA gain for all driver electrons is $\sim200$MeV. Owing to the DLA, electrons stay in the decelerating (front) portion of the bubble for a much longer time than they would have stayed without the laser pulse. In fact, in the absence of the DLA, the driver bunch would have lost most of its energy after less than $z \approx 10$mm of propagation through the plasma due to rapid self-deceleration by its own wakefield. The physics underlying the extended bunch propagation during Stage 1 of LEPA is described in detail in Sec.~\ref{sec:extended_bunch_prop}.

The second stage of LEPA (from $z=z_1$ to $z_2=28$mm for this specific example) is characterized by the bunch electrons getting out of resonance with the laser pulse. The physics of falling out of resonance has been described elsewhere~\cite{Farfield}, and will not be described here. During Stage 2, driver electrons either get decelerated by the wakefield and slip back into the back portion of the bubble, or move out of the bubble entirely because of the increased amplitude of their betatron oscillations~\cite{Farfield}.  As shown in Fig.~2~(c) and (g), the majority of driver bunch electrons have already slipped back into the decelerating portion of the wake at $z=z_2$. This has two effects, both of which are deleterious to the acceleration of a witness beam. First, the plasma bubble becomes smaller because it is primarily driven by the laser pulse. Second, the wake is actually depleted via the beam-loading effect produced by the driver bunch in the back portion of the plasma bubble.

The third stage of LEPA, during which the laser pulse continues to propagating in a self-guided regime until its complete depletion, takes place from $z=z_2$ to $z_3=62$mm: see Figs.~\ref{Fig2}~(d, h). The witness beam experiences the highest average acceleration gradient of the order $E_{\parallel} \approx 90 {\rm GV/m}$ during the first two stages ($z_0 < z < z_2$) because the driver bunch enhances the bubble created by the laser pulse. On the other hand, a smaller average acceleration gradient of approximately $E_{\parallel} \approx 58$GV/m is experienced by the witness beam during the third stage of LEPA ($z_2 < z < z_3$). Overall, the witness beam gains $W_{\rm LEPA} \approx 4.5$GeV of energy in total over a distance of $L_{\rm LEPA} = z_3  \approx 62$mm.

\subsection{Evidence of synergy between electron bunch and laser pulse}

Understanding whether the above energy gain of $W_{\rm LEPA}$ is sufficient for justifying the hypothesis of synergy between the bunch-guided laser pulse and laser-accelerated bunch requires a quantitative comparison between energy gains by a witness bunch when identical laser pulse and electron driver are used separately. For example, one can envision two sequential plasma accelerator stages: a LWFA driven by a laser pulse alone, followed by a PWFA driven by an electron bunch; the laser, beam, and plasma parameters for both stages are listed in Sec.~\ref{sec:params}. The results of these simulations are listed in Table I.

In the case of the same plasma density for LWFA and PWFA listed in Sec.~\ref{sec:params}, and a low-charge accelerated witness beam, the results are listed in the first three columns, top row of Table I. Laser-beam synergy is confirmed by observing that $W_{\rm LEPA} > W_{\rm LWFA} + W_{\rm PWFA}$ in the case when the depletion (also known as loading) of the plasma wake by the witness bunch can be neglected. This result is quite remarkable given that the driver bunch can significantly deplete the plasma wake after most of its electrons slip into the back (accelerating) portion of the plasma bubble. The synergy between the driver bunch and the laser pulse is even more pronounced in the case of a moderately-charged witness beam with a total charge of $q=0.18$nC (bottom row of Table I): the energy gain in a LEPA scheme exceeds the sum of energy gains in sequential LWFA and PWFA schemes by over $60\%$ while producing smaller energy spreads.

\begin{table}[ht]
\caption{\label{table1}Energy Gain $W$ and Emittance for Different Acceleration Scheme}
\begin{ruledtabular}
\begin{tabular}{lcccc}
\
Scheme&LEPA&PWFA&LWFA&LWFA-OPT\footnote{Laser-only with optimal parameters.}\\
\hline
Gain\footnote{No beam loading.}(GeV)         &4.5&1.3&2.4&3.0\\
Gain\footnote{Beam loading = 0.18nC.}(GeV)   &$3.6\pm0.25$&$1.0\pm0.25$&$1.2\pm0.4$&$1.5\pm0.7$\\
Emittance$^{c,}$\footnote{unit = $mm\cdot mrad$} &$1.7\times10^{-3}$&$1.7\times10^{-3}$&$1.0\times10^{-3}$&$0.5\times10^{-3}$\\
\end{tabular}
\end{ruledtabular}
\end{table}

We note that while $W_{\rm PWFA}$ is limited by the transformer ratio, it is possible to increase the energy gain of a LWFA by keeping $U^{\rm laser}$ the same while reducing the pulse duration and increasing the plasma density in the LWFA stage (see Sec.~\ref{sec:comparison} for details). The values of thus optimized $W_{\rm LWFA}$ without (with) witness beam loading are listed in top (bottom) rows, the fourth column of Table I. Such optimization of the LWFA stage does not change our conclusion: the synergy between the driver beam and the laser pulse enables an overall increase of the energy gained by the witness bunch: $W_{\rm LEPA} > W_{\rm LWFA} + W_{\rm PWFA}$ with and without beam loading. The key to understanding such synergy lies in analyzing the physics responsible for the extension of the laser-bunch propagation in LEPA described in the following Sec.~\ref{sec:extended_bunch_prop}. Specifically, we demonstrate the extension (i) of the driver beam propagation by the DLA mechanism (see Sec.~\ref{sec:DLA}), and (ii) of the laser pulse propagation by the bunch channeling (see Sec.~\ref{sec:beam_channel}) during Stage 1 of LEPA. Further details of the comparison between LEPA, LWFA, and PWFA scheme are presented in Sec.~\ref{sec:comparison}.

\section{Physics of the Synergistic Laser-Bunch Propagation}\label{sec:extended_bunch_prop}
Assuming that the electrons in the driver bunch start out with the initial longitudinal momentum $p_{b} = \sqrt{\gamma_b^2 - 1} mc$, and that the average decelerating wakefield across the whole bunch is $\bar{E}_{\parallel}$, we can estimate the maximum distance $L_{\rm dec}$ traveled by a typical electron without any additional energy input from the laser: $L_{\rm dec} \approx \gamma_b mc^2/e\bar{E}_{\parallel}$. For the example illustrated by Fig.~\ref{Fig2} (see caption for laser, plasma, and electron bunch parameters), we estimate that $L_{\rm dec} \approx 10$mm. A very similar estimate is obtained by assuming that the decelerating wakefield is of the same order as the cold plasma wavebreaking field $E_{\rm WB} \equiv mc\omega_p/e \approx \sqrt{n_0/10^{18}{\rm cm^{-3}}} [{\rm GV/cm}]]$.

\begin{figure}[htp!]
\centering
  \includegraphics[width=0.98\columnwidth]{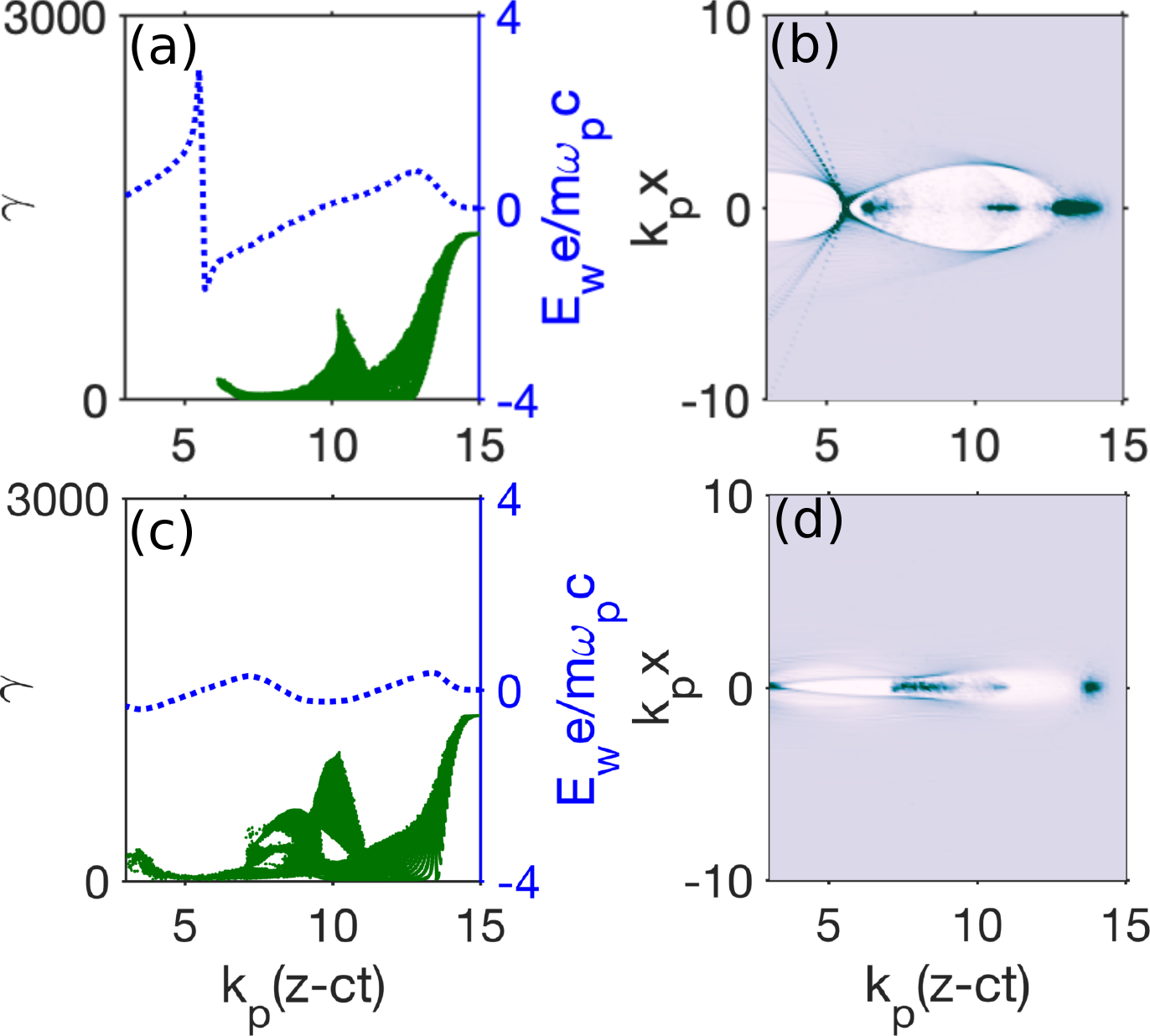}\\
\caption{Simulation of the driver bunch dynamics in a PWFA at (a,b) $z=z_1 = 12$mm and (c,d) $z=20$mm. (a,c) Longitudinal phase space $(k_p \xi, \gamma)$ of the driver bunch electrons (dark green dots) and wakefield $E_{\parallel}=E_z$ (blue dashed line). (b,d) Plasma density profile and the plasma bubble generated by the driver bunch. Electron bunch and plasma parameters: same as in Fig.~\ref{Fig2}.}\label{Fig3}
\end{figure}

These estimates have been validated with a bunch-only (i.e PWFA: no laser pulse) WAND-PIC simulation. The results are presented in Fig.~\ref{Fig3}. Because of the short duration of the bunch, $\tau_b \sim 2\omega_p^{-1}$, all driver electrons are decelerated after a short propagation distance of $z=z_1$, as evident from Fig.~\ref{Fig3}~(a). Further deceleration and spreading of the driver bunch results in a continuous decrease of the accelerating gradient shown as a dashed line in Fig.~\ref{Fig3}~(c) corresponding to $z=20$mm propagation distance. This is accompanied by even further reduction of the bubble size as shown in Fig.~\ref{Fig3}~(d). Clearly, the driver bunch no longer supports a robust bubble structure at $z=20$mm. Additionally, the bubble size reduction in a PWFA adversely affects the acceleration of a witness beam placed in the back of the original ($z=0$) plasma bubble. In agreement with the transformer ratio limit, the maximum energy gain of the witness beam is $\Delta W_{\rm PBWA} = 2\gamma_b mc^2 \approx 1.3$GeV (assuming infinitesimal witness charge).

The comparison between Fig.~\ref{Fig3}~(a) and Fig.~\ref{Fig2}~(f) demonstrates that for the same propagation distance $z=z_1$, all of the driver electrons in the bunch-only case are decelerated while many of the driver bunch electrons at the front and rear of the bubble are actually accelerated in LEPA. Crucially, those electrons in the front (decelerating) portion of the bubble are accelerated by the DLA mechanism. Maintaining electrons in the front of the bubble is crucial for maintaining the size and the accelerating field of the bubble: both are larger in the LEPA scenario (see Fig.~\ref{Fig2}~(b)) than in the standard PWFA (see Fig.~\ref{Fig3}~(b)) scenario. Next, we investigate the role of DLA in extending the propagation distance of the driver bunch.

\subsection{Synergy of LEPA: DLA-extended electron bunch propagation}\label{sec:DLA}

The DLA mechanism can counter the decelerating field and, in fact, accelerate bunch electrons at a rate of $d\gamma/dt \approx e\bar{E}_{\parallel}/mc^2$. The maximum energy achieved by the electron while in resonance with the laser field has been estimated as $\epsilon_{\rm max} \approx mc^2(\omega_p/\omega_0)^2( E_0/\bar{E}_{\parallel})^4/130$~\cite{Farfield}. The approximate equation for betatron resonance of a laser pulse with an electron with the energy $\epsilon_{\rm res} \equiv \gamma_{\rm res} mc^2$ undergoing a transverse undulation with the betatron frequency $\omega_{\beta} \approx \omega_p/\sqrt{ 2\gamma_{\rm res}}$ and time-averaged transverse momentum $p_{\perp}$ is given by~\cite{Zhang_PPCF16}
\begin{equation}\label{eq:resonance}
  \omega_{\beta}(\gamma_{\rm res}) = \omega_0\left(\frac{1+p_{\perp}^2/m^2c^2}{2\gamma_{\rm res}^2}+\frac{1}{2\gamma_{\rm ph}^2}\right),
\end{equation}
where the relationship between the laser wavenumber $k_0$ and frequency $\omega_0$ is expressed as $\omega_0^2 = c^2k_0^2 \left( 1 + \gamma_{\rm ph}^{-2} \right)$.

A typical driver electron with a large initial momentum $\gamma_b \gg \gamma_{\rm res}$ undergoes the following sequence of energy exchanges with the wake and the laser field. Initially, most electrons are decelerated by the wakefield to the energy comparable to $\epsilon_{\rm res}$ because of the resonance condition given by Eq.(\ref{eq:resonance}) is not initially satisfied. Such deceleration takes place during the $0 < z < z_{\rm dec}$ interval, where $z_{\rm dec} \approx 8$mm. Note that $z_{\rm dec} \ll L_{\rm dec}$ because the bunch is decelerated by the combined wakefield: its own, and that of the laser pulse.

Next, most of the driver bunch electrons become resonant with the laser field and start gaining energy from the DLA process. As can be observed in Fig.~\ref{Fig2}~(f), at least half of the electrons regain most of their energy from the DLA process and stay in the front portion of the plasma bubble. Finally, as resonant electrons gain significant transverse and total energy, the resonant condition is no longer be satisfied. The total distance that traveled by a typical resonant electron before detuning away from the DLA resonance can be estimated as $L_{\rm DLA} \approx z_{\rm dec} + mc^2(\omega_p/\omega_0)^2 E_0^4/\left( 130 e \bar{E}_{\parallel}^5 \right)$. Assuming that $e\bar{E}_{\parallel}/m\omega_pc \sim 1$ ~\cite{Lu_GeV,Pukhov_Pheno_2004}), the propagation distance of the DLA-assisted driver is longer than the self-stopping distance by the following factor: $L_{\rm DLA}/L_{\rm dec} \approx 1 + (mc/\gamma_b) (\omega_0/\omega_p)^2 a_0^4/130$. For the parameters listed in the caption of Fig.~\ref{Fig2}, we find that $L_{\rm DLA} \approx 2.7 L_{\rm dec}$. Therefore, we estimate that $L_{\rm DLA} \approx 21.6$mm. This distance is in good agreement with the value of $z_2$ observed in Fig.~\ref{Fig2}~(g).

\begin{figure}[htp!]
\centering
  \includegraphics[width=0.99\columnwidth]{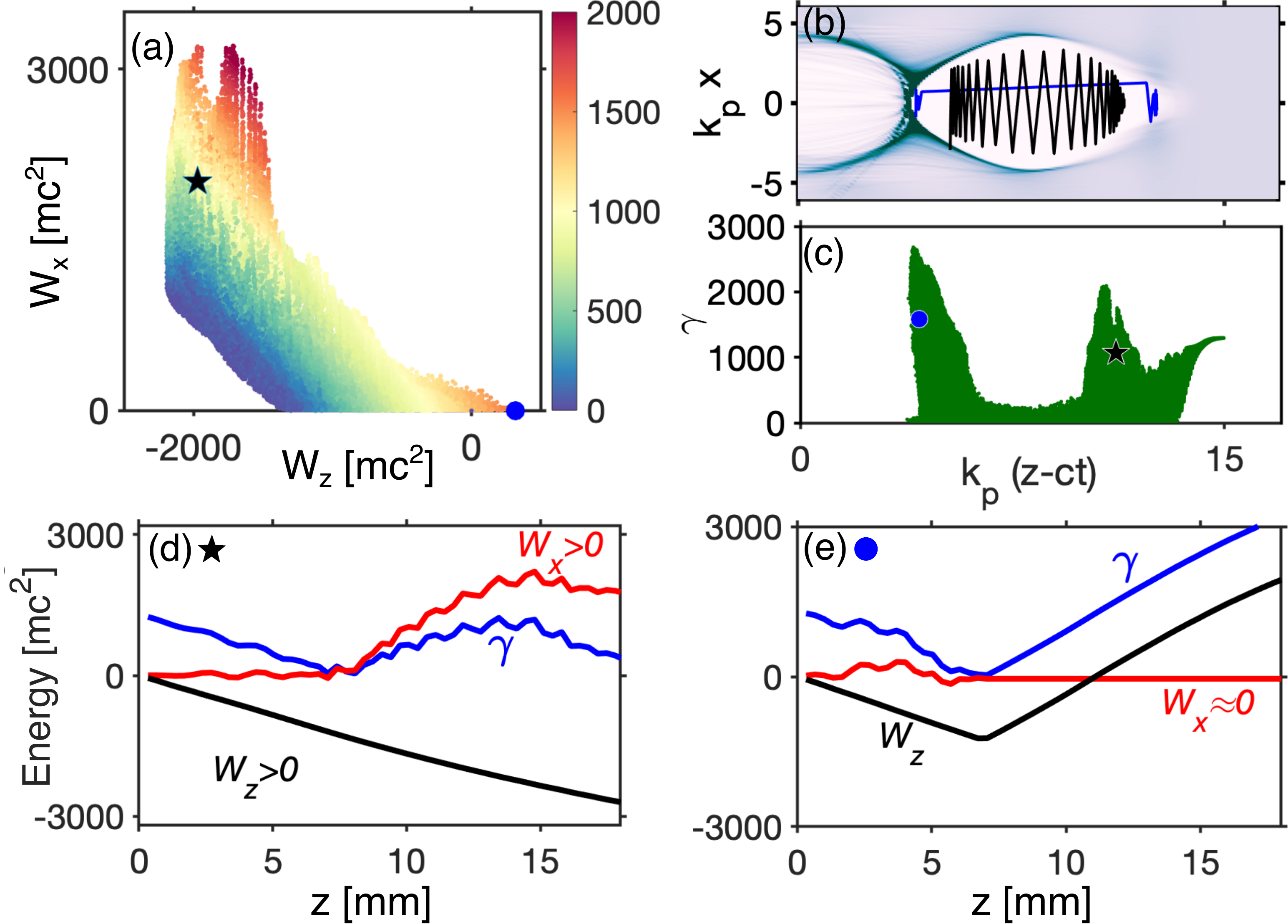}\\
\caption{Electron bunch interaction with the laser and wakefield, and the dynamics of two representative resonant and non-resonant electrons. (a) Work done by the laser ($W_x$) and wakefield ($W_z$) on all electrons of the driver bunch color-coded by their final relativistic factor $\gamma$. (b) Trajectories of the resonant (black line) and non-resonant (blue line) electrons in the plasma bubble.  (c) Longitudinal phase space $(\xi,\gamma)$ of the bunch electrons. (d,e) Time evolution of the laser work $W_x$ (red), $\gamma$ (blue), and wakefield work $W_z$ (black) of a resonant (d) and a non-resonant (e) electrons. Resonant and non-resonant electrons are, respecttively, marked by a black star and blue circle in (a,c,d,e). Propagation distance in (a,c): $z=12$mm.}\label{Fig4}
\end{figure}

To further clarify the role of DLA in extending driver bunch propagation in LEPA, it is instructive to separately calculate the work done by the laser and by the wakefield on a given $j$'th electron: $W_x^{(j)}$ and $W_z^{(j)}$, respectively. The dimensionless total energy of each electron, $\gamma^{(j)} = \gamma_b + \left( W_x^{(j)} + W_z^{(j)}\right)/mc^2$, is color-coded in the $(W_x,W_z)$ space and plotted in Fig.~\ref{Fig4}~(a) for all driver bunch electrons at the propagation distance $z=z_1$ through the plasma. Fig.~\ref{Fig4}~(a) is convenient for classifying the electron population into two distinct resonant ("DLA") and non-resonant ("non-DLA") sub-populations. For convenience, we classify those with $W_x^{(j)}>200mc^2$ as DLA electrons (approximately $50\%$ of all driver electrons), and the remaining as non-DLA electrons (the superscript $j$ is dropped hereafter).

As we see from Fig.~\ref{Fig4}~(a), the DLA sub-population electrons have gained energy up to $W_x \approx 3,000mc^2$ directly from the electric field of the laser; the average DLA energy gain for those electrons is $\langle W_x \rangle \approx 860mc^2$. We further observe from Fig.~\ref{Fig4}~(a) that those DLA electrons which gained more energy from the laser field also lost more energy to wakefield, i.e. $W_x$ and $W_z$ are anti-correlated. This implies that the electrons with the largest $W_x$ stayed in the decelerating (front) region of the bubble for a longer time. Therefore, resonant DLA extended their propagation distance in the front portion of the bubble and potentially contributed to enhancing the wakefield strength and the size of the plasma bubble. On the other hand, the non-DLA electrons exhibit much smaller (yet still negative) energy exchange with the wakefield. This implies that they have rapidly slipped to the back of the bubble, where their positive energy gain from the wake partially offset their initial energy loss to the wake.

To further investigate the difference between DLA and non-DLA sub-populations, it is helpful to select one representative DLA (black star) and one non-DLA (blue circle) electron marked in Fig.~4~(a). The trajectories of the DLA and non-DLA electrons (black and blue lines, respectively) inside the plasma bubble are plotted in Fig.~\ref{Fig4}~(b). Although the non-DLA electron starts its motion at the head of the driver bunch, where the decelerating field is smaller, it stays in the decelerating field for a much shorter time compared to the DLA electron. For example, at $z=z_1$ the non-DLA electron has already slipped to the back of the bubble, whereas the DLA electron remains in the decelerating field. The energies and longitudinal positions of these two electrons inside are shown in Fig.~4~(c) as a blue circle and a black star.

The evolution of $W_x(z)$, $W_z(z)$, and $\gamma(z)$ for the representative DLA electron is plotted in Fig.~4~(d) as a function of the propagation distance. From $z=0$mm to $z=z_1$, the DLA electron gains $W_x = 1.0$GeV from the laser pulse and loses $W_z = -1.05$GeV to the wakefield. Note that the gain $W_x>0$ only starts at $z=7.0$mm illustrating that this electron indeed gains energy from the laser at nearly twice the rate of its loss. Its further trajectory illustrates a rather typical fate of a DLA electron: it keeps propagating in the decelerating field of the bubble up to $z=26$mm, and then leaves the bubble. In contrast, its non-DLA counterpart in Fig.~4~(e) stays in the decelerating field of the bubble for up to $z=7.5$mm. This comparison indicates that the propagation distance of the DLA electron got in the leading portion of the plasma bubble is extended by $\Delta z = 18$mm, in a good agreement with the earlier provided estimate. Having demonstrated that the DLA mechanism extends the electron bunch propagation, we now demonstrate how the presence of the electron bunch extends the propagation of the laser pulse itself.

\subsection{Synergy of LEPA: Laser Pulse Guiding by the Driver Bunch}\label{sec:beam_channel}

\begin{figure}[htp!]
\centering
  \includegraphics[width=0.99\columnwidth]{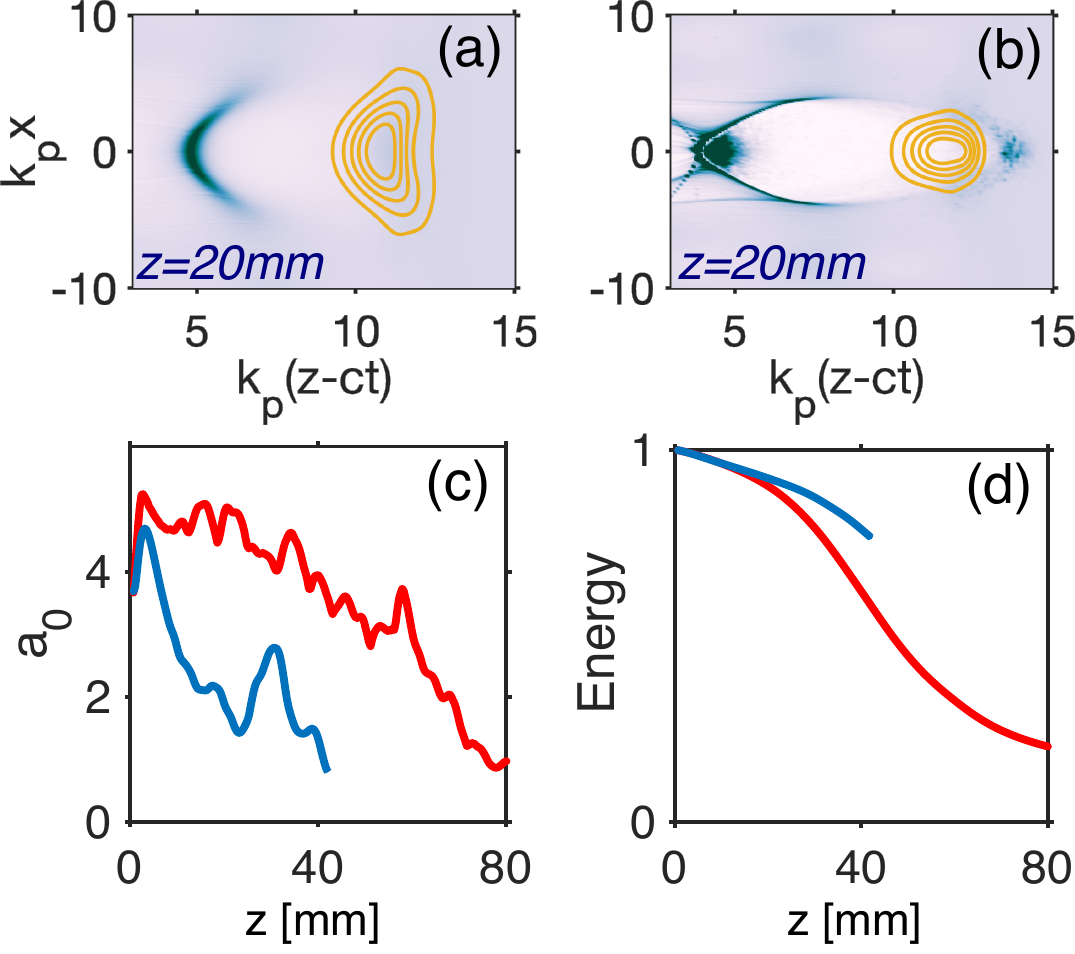}\\
\caption{Comparison between LWFA and LEPA scenarios.  Plasma bubble driven by laser pulse without (a) and with (b) the driver bunch, both at $z=20$mm. Yellow lines: laser intensity isocontours. (c,d) Evolution of the peak vector potential $a_0$ (c) and the total energy (d) of the laser pulse for the LWFA (blue) and LEPA (red) scenarios.}
\label{Fig5}
\end{figure}

As the driver bunch gains energy and increases its propagation distance due to its DLA interaction with the laser pulse, it concurrently extends the propagation distance of the laser pulse. Specifically, during the first and second stages of LEPA, the bunch can sustain a deeper plasma channel that channels the laser pulse and mitigates laser diffraction~\cite{BeamChannel_Shvets_PhysRevE97}. The channeling effect takes place in addition to the enlargement of the laser-produced plasma bubble by the bunch, and of the corresponding accelerating field at the back of the bubble.

To illustrate the channeling effect, we have compared laser propagation through the plasma with (LEPA) and without (LWFA) the driver bunch. The results are presented in Fig.~\ref{Fig5}. By comparing Figs.~\ref{Fig5} (a) and (b), we observe that the spot size of the bunch-guided laser pulse in the LEPA scenario is considerably smaller than that of the self-guided laser pulse in the standard LWFA scheme. The diffraction of the laser pulse in the LWFA scheme results in a significant drop (by almost a factor $2$) of its peak on-axis normalized vector potential $a_0$. This should be contrasted with almost constant $a_0$ in the LEPA configuration, as shown in Fig.~\ref{Fig5} (c). In fact, the laser has essentially diffracted by the time it propagated $z=40$mm into the plasma, and the peak $a_0$ drops below unity. In contrast, the laser pulse propagates up to $z=z_3 \approx 62$mm in LEPA.

Relatively rapid diffraction of the self-guided laser owes to its moderate power: $P/P_{\rm crit} \approx 5$. Therefore, the maximum energy gain of a low-charge witness beam is only $W_{\rm LWFA} \approx 2.4$GeV. On the other hand, the laser pulse in LEPA is channeled by the bunch, thereby gaining an additional $\Delta z = 24$mm of propagation while maintaining a much higher intensity (see Fig.~\ref{Fig5} (c) for comparison). Crucially for the efficiency of a LWFA, the self-guided laser pulse diffracts before it loses just $\eta_{\rm LWFA} \approx 20\%$ of its total energy. This happens because a shallow plasma bubble created by a transversely-spread laser pulse is a poor absorber of the laser energy~\cite{Mora_1997}. On the other hand, the bunch-driven laser pulse transfers $\eta_{\rm LEPA} \approx 70\%$ of its energy to the plasma, thereby creating a larger accelerating gradient and providing a larger energy gain of $\Delta W_{\rm LEPA} \approx 4.5$GeV to a low-charge witness beam.

Figure~\ref{Fig5}~(d) presents the comparison between laser energy absorption by the plasma in the LWFA (blue line) and LEPA (red line) schemes as a function of the propagation distance. The higher depletion of the laser pulse in the LEPA scheme explains the higher energy gain $W_{\rm LEPA}$ in comparison with the energy gain $W_{\rm LWFA}$ in a LWFA. Note that the ratio $W_{\rm LEPA}/W_{\rm LWFA} $of the gained energies in the LEPA and LWFA schemes is considerably smaller than the ratio $\eta_{\rm LEPA}/\eta_{\rm LWFA}$ of the extracted laser energies. Plasma wake depletion ("loading") by the driver bunch electrons that eventually slip into the accelerating portion of the bubble account for this discrepancy.

\section{Comparison Between Different Schemes: Beam Loading and Optimization}\label{sec:comparison}
The results of our numerical simulations listed in Table~I show that the combination of a laser pulse and a driver bunch result in considerably larger energy gain when compared to the laser-only or bunch-only scenarios. Moreover, the energy gain in LEPA is super-additive: $W_{\rm LEPA} > W_{\rm LWFA} + W_{\rm PWFA}$. One part of the enhanced energy gain comes from the fact that the channel created by the bunch enhances the focusing and guiding of the laser pulse at the first stage of LEPA. Surprisingly, this effect lasts even past the driver bunch depletion. The reason is that the wavefront of the laser pulse is re-adjusted and flattened in the deep channel created by the bunch as can be observed by comparing Figs.~\ref{Fig5}~(a) and (b). Such a laser pulse profile is beneficial to transversely-confined propagation even after the guiding bunch slips back from the front of the bubble. Another part of the enhanced energy gain comes from the DLA effect which extends the bunch propagation distance.

\subsection{Plasma Wake Loading by A Witness Bunch}\label{sec:beam_loading}
Because additional driver energy ($\sim3\%$ of the laser energy) is expended in the LEPA scheme, it is instructive to investigate how much additional energy can be imparted to a witness beam with finite charge $q$. We have carried out numerical simulations with WAND-PIC for all three acceleration schemes and included the beam loading of the wake by the witness beam. Note that beam loading by the driver bunch electrons that eventually slip to the back of the bubble has been already accounted for in the simulations presented in Figs.\ref{Fig2}, \ref{Fig3}, \ref{Fig4}. Therefore, in what follows we drop the direct reference to the witness beam when discussing beam loading.

In the anticipation that a significant portion of the driver bunch energy will be utilized to increase the energy content of the witness bunch, we have selected the witness bunch charge according to $q=Q_b \left(\gamma_b mc^2/W_{\rm LEPA} \right)$. Gaussian witness beam's transverse and longitudinal sizes were chosen to be $w_x = w_y = 4.2\mu$m and $w_z=2.8\mu$m, respectively. Note that the longitudinal density profile of a witness bunch can be optimized to reduce the energy spread~\cite{Witness_Shaping,Witness_Shaping_2}. Such optimization is beyond the scope of this work and will be the subject of future research. Throughout this paper, Gaussian density profiles are assumed for both driver and witness bunches.   

The results of the beam-loaded simulations are listed in the second row of Table.~I.  The average energy gain of the witness bunch in the LEPA scheme is $\langle W_{\rm LEPA} \rangle \approx 3.6$GeV, with a $\pm 0.25$GeV energy spread due to beam loading. The reduction from $W_{\rm LEPA} \approx 4.5$GeV is also due to beam loading. The average energy gain in the PWFA is also reduced: $\langle W_{\rm PWFA} \rangle \approx 1$ GeV. The largest energy gain reduction is found in the LWFA case: $\langle W_{\rm LWFA} \rangle \approx 1.2$GeV, which is only $50\%$ of the $W_{\rm LWFA} \approx 2.4$GeV energy gain in the absence of beam loading. We also listed the transverse emittance of the wittiness beam in the bottom row of Table.~I. The witness beams in LEPA, PWFA, and PWFA schemes have comparable emittance. 

This observation reveals another advantage in combining the laser pulse and an electron bunch in a LEPA: a steep bubble created by the two is less susceptible to depletion by a witness beam. On the other hand, the bubble created by the laser alone in a LWFA at moderate power is relatively shallow, and thus more susceptible to beam loading.  Overall, the witness beam gains $\Delta W \equiv \left (\langle W_{\rm LEPA} \rangle - \langle W_{\rm LWFA} \rangle \right) \approx 2.4$GeV more energy per electron in the LEPA than in the LWFA scenario. Therefore, the excess energy gain of the witness bunch is $\Delta U_{\rm witt} = q \Delta W \approx 0.4$J. Remarkably, $\Delta U_{\rm witt}$ is only a factor $2$ smaller than $U^{\rm bunch} \approx 0.8{\rm J}$. This confirms our initial expectation of high energy utilization of the driver bunch by the witness beam. 

\begin{figure}[htp!]
\centering
  \includegraphics[width=0.99\columnwidth]{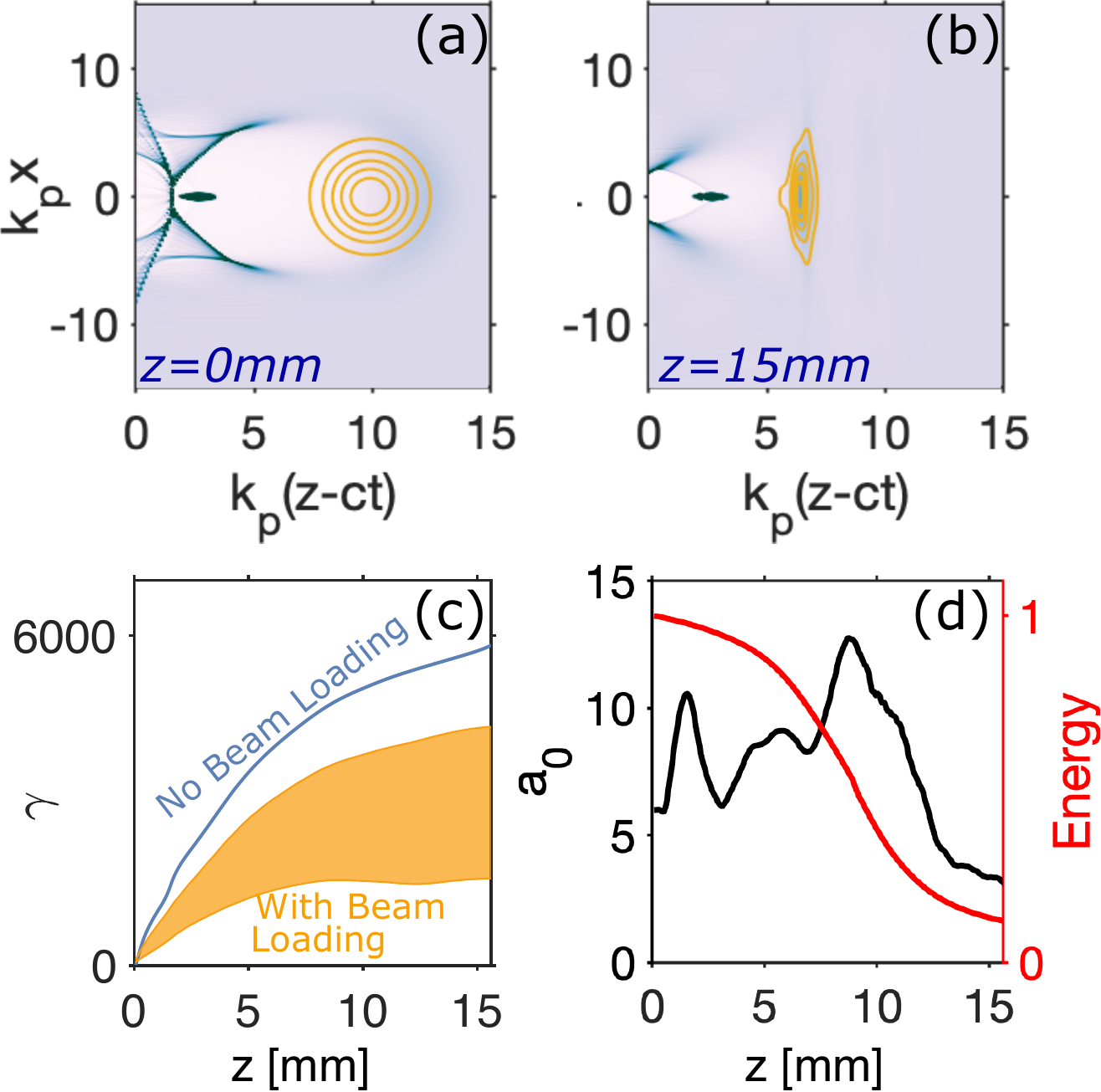}\\
\caption{Simulation of an optimized LWFA. (a) Plasma bubble at the initial moment. (b) Plasma bubble near laser depletion. (c) Evolution of the energies of witness beam  when beam loading is switched on and off. (d) The evolution of vector potential $a_0$: black line and laser energy: red line.}
\label{Fig6}
\end{figure}

In comparing energy gains of the witness bunch in the LEPA {\it vis a vis} LWFA and PWFA schemes, we have assumed that the same plasma density $n_0=4\times10^{17}$ cm$^{-3}$ is used for all three schemes. In fact, it is reasonable to ask if the same laser energy $U^{\rm laser} \approx 27 {\rm J}$ can be deployed more efficiently is a LWFA scheme by shortening the pulse and increasing the plasma density. Such an approach is based on increasing the ratio of the peak laser power $P$ (which increases inverse proportionally to laser pulse duration $\tau_{\rm FWHM}$) to critical laser power $P_c$ (which decreases proportionally to plasma density). A higher $P/P_c$ ratio contributes to longer laser pulse propagation length and overall efficiency of laser energy deposition into the plasma. Therefore, we have identified
a set of "optimal" laser-plasma parameters that result in the longest laser propagation and the largest energy gain of the witness beam. We refer to this scheme as LWFA-OPT.

Specifically, the plasma density was increased to $n_0^{\rm OPT} = 1.5\times10^{18}$cm$^{-3}$ and the laser pulse duration was reduced to $\tau_{\rm FWHM}^{\rm OPT}=49$fs, thereby increasing the laser power to $P=533 {\rm TW} \approx 27P_c$ while preserving its total energy. The normalized vector potential $a_0=6$ is chosen based on the optimized matching conditions~\cite{Lu_GeV}. The laser duration is chosen to have a comparable dephasing and depletion lengths~\cite{Lu_GeV}. The results of the WAND-PIC simulation are plotted in Fig.~\ref{Fig6}. As one can observe from Fig.~\ref{Fig6}~(b), the laser pulse propagates up to $z=15$mm, at which point the depletion of the pulse and the dephasing of the witness beam were simultaneously achieved. Without the beam loading, the maximum possible gain of a witness beam is $W_{\rm LWFA}^{\rm OPT} \sim 3$GeV. For a realistic witness beam with a moderate charge $q=0.18$nC charge, the average energy gain is reduced to $\langle W_{\rm LWFA}^{\rm OPT} \rangle \sim 1.5$GeV, with the energy spread of $ \Delta W = \pm 0.7$GeV. Therefore, the LEPA scheme still enables larger energy gain ($\langle W_{\rm LEPA} \rangle > \langle W_{\rm LWFA}^{\rm OPT} \rangle + \langle W_{\rm PWFA} \rangle$) and smaller energy spread.

\section{Discussion and Conclusions}\label{sec:conclusion}

LEPA appears to be a promising plasma-based acceleration scheme, and an effective way of producing Multi-GeV single-stage energy gain of a witness bunch using moderate energy ($U^{\rm laser} \approx 27 {\rm J}$) laser pulses and high-charge ($Q_b=1.25$nC) relativistic ($\gamma_b mc^2 = 650$MeV) driver bunches. When comparing LEPA to the more conventional plasma-based accelerator approaches, such as LWFA and PWFA, we have discovered that the propagation distances of both the electron bunch and the laser pulse are extended as the result of a synergistic interaction between the laser pulse and the bunch. LEPA is found to be particularly promising in the regime of significant depletion of the plasma wake by a moderately-charged ($q=180$pC) witness bunch, i.e., a $140\%$ more energy gain is achieved compared with the optimized LWFA. Moreover, combining a diver bunch with the laser pulse doesn't worsen the beam quality, in fact, the LEPA can achieve less energy spread than LWFA and comparable beam emittance when compared with both PWFA and LWFA. The scheme does not require an external plasma channel for laser guiding, however, good alignment of the driver bunch and laser pulse is required to drive the bubble steadily over centimeters. Further simulations indicate that for the main LEPA simulation we showed in the paper, a miss-alignment angle $<10^{-3}$ is required. We can envision using external linacs for providing driver bunches for each acceleration stage. Using a low-quality electron bunch from another laser-plasma accelerator would also be an option for conducting early proof-of-principle experiments without major investments in acceleration infrastructure.

While the driver bunch plays mostly an auxiliary role in the LEPA approach, such as guiding the more energy-rich laser pulse and increasing the size of the accelerating plasma bubble, we speculate that the driver bunch does not need to be totally "wasted" after each LEPA stage. As we can clearly observe from Fig.~\ref{Fig2}~(h), many of the driver bunch electrons have gained significant energy from the wake and from the laser via DLA. In our example, at distance $z=40$mm, a considerable charge of $Q_1 \approx 0.84$ nC achieves total energies above $1$ GeV, and they form a new bunch with transverse size $=9{\rm \mu m}$ and duration $=28fs$. Such a bunch can be reused for an additional stage of a PWFA. Moreover, considerable transverse energy is acquired by the driver bunch electrons during its DLA. In our example, at least $55\%$ $(N = 4.3\times10^{9}$) electrons acquire transverse momenta larger than $p_{\perp} = 10mc$. These electrons can be used for x-ray or even $\gamma$-ray generation. Moreover, such radiation can be used as the diagnostic of the individual acceleration stages of LEPA.

Several factors are limiting the performance of the LEPA scheme. One is the DLA performance: DLA in decelerating wakefield increases the transverse momentum of the driver electrons~\cite{Farfield} and the DLA distance is limited by the resonance condition which cannot be met by all electrons. Not every driver electron can reach the maximum propagation distance before exiting the bubble or the resonance. This could be improved by careful engineering of the driver electrons' phase space. The second issue is the beam loading by the driver electrons. Non-DLA and some of the DLA electrons slip to the back of the bubble and reduce the accelerating gradient. This naturally reduces the final energy gain of the witness beam and introduces additional challenges to limiting the energy spread of the witness beam. We speculate that under some conditions such wake depletion by the driver bunch can be reduced by utilizing the bubble contraction. Overall, synergistic interactions between co-propagating electron bunches and laser pulses open up new physical effects that are likely to be explored for a variety of acceleration and radiation generation applications.

\section{Acknowledgments}
This work was supported by the DOE grant DE-SC0019431. The authors thank the Texas Advanced Computing Center (TACC) at The University of Texas at Austin for providing HPC resources.

\end{document}